\documentstyle{article}

\font\tenrm=cmr10
\font\tenit=cmti10
\font\elevenbf=cmbx10 scaled\magstep 1
\font\elevenrm=cmr10 scaled\magstep 1
\font\elevenit=cmti10 scaled\magstep 1

\font\ninerm=cmr9

\renewenvironment{thebibliography}[1]
 { \elevenrm
   \begin{list}{\arabic{enumi}.}
    {\usecounter{enumi} \setlength{\parsep}{0pt}
     \setlength{\itemsep}{3pt} \settowidth{\labelwidth}{#1.}
     \sloppy
    }}{\end{list}}

\begin{document}
\begin{center}
\vglue 0.6cm
{%{\tenbf WORLD SCIENTIFIC PUBLISHING COMPANY\\}
 {\elevenbf        \vglue 10pt
               Symmetries and Symmetry-Breaking\\
               \vglue 3pt
               in String Theory\footnote{
\ninerm\baselineskip=11pt Summary of a talk given at SUSY93,
Northeastern University, April 1, 1993.
}
\\}
\vglue 5pt
%{\ninerm (For 20\% Reduction to 8.5 $\times$ 6 in Trim Size)\\}
\vglue 1.0cm
{\tenrm Gregory Moore \\}
\baselineskip=13pt
{\tenit  Department of Physics, Yale University, \\}
\baselineskip=12pt
{\tenit New Haven, CT 06511, USA\\}}

%\vglue 0.3cm
%{\tenrm and\\}
%\vglue 0.3cm
%{\tenrm SECOND AUTHOR'S NAME\\}
%{\tenit Group, Company, Address, City, State ZIP/Zone, Country\\}
%\vglue 0.8cm
%{\tenrm ABSTRACT}

\end{center}

%\vglue 0.3cm
%{\rightskip=3pc
% \leftskip=3pc
% \tenrm\baselineskip=12pt
% \noindent
%Typeset the abstract with an indentation
%of 3 picas on the left and right margins and in 10 point roman with
%baselineskip of 12 point.}
\def\IL{\relax{\rm I\kern-.18em L}}
\def\IR{\relax{\rm I\kern-.18em R}}
\def\CM {{\cal M}}
\def\CN {{\cal N}}
\def\CR {{\cal R}}
\def\CD {{\cal D}}
\def\CF {{\cal F}}
\def\CP {{\cal P }}
\def\CL {{\cal L}}
\def\CV {{\cal V}}
\def\CO {{\cal O}}
\def\CZ {{\cal Z}}
\def\CE {{\cal E}}
\def\CH {{\cal H}}
\def\CC {{\cal C}}
\def\CA {{\cal A}}
\def\IL{\relax{\rm I\kern-.18em L}}
\def\IR{\relax{\rm I\kern-.18em R}}
\def\IH{\relax{\rm I\kern-.18em H}}
\def \CC{{\cal C}}
\def \p{\partial}
\def\pb{\bar{\partial}}

\vglue 0.6cm
{\elevenbf\noindent 1. Introduction}
\vglue 0.2cm
{\elevenit\noindent 1.1. Motivations}
\vglue 0.1cm
\baselineskip=14pt
\elevenrm
This note summarizes a talk based on
\cite{finite}.  One motivation for this work
is an old and basic question in string theory: What are the
fundamental symmetry principles upon which theory should be
based. If we understood the answer to this we would
then go on to ask how the (on shell) symmetries are
broken, and whether there is a notion of a
``most symmetric background.''
In asking these questions we have in mind
analogous systems - Yang-Mills theories and general
relativity. In the Yang-Mills example the answers
to the above questions are known: the fundamental symmetry
principles are local gauge invariance and Poincar\'e invariance.
Symmetries can be broken when a Higgs scalar $\phi$ takes a noninvariant
expectation value. A symmetric background leaving the symmetries
unbroken is $\langle \phi\rangle = 0$.

A second set of motivations comes from recent progress on
two fronts in string theory: First, in the past few years a
covariant closed string field theory (CSFT) has finally been
constructed \cite{zwiebach}.  Unfortunately, the current formulation
of the theory is rather complicated. One naturally
hopes that a deeper understanding of the symmetry principles
underlying CSFT will lead to simplification. A second piece
of progress is one of the fruits of the
matrix-model/2d gravity developments of 1989-1992.
This was the realization
that certain two-dimensional string backgrounds have large
unbroken symmetries. Roughly speaking, they have a $W_\infty$ symmetry
\cite{wittzwie}.

By a symmetry of the background we mean the following \cite{wittzwie}.
In CSFT a CFT $\CC$ of $c=26$ is a classical solution to the equations
of motion. The string field $\Psi$ represents deviations of
the fields from this solution.
CSFT is a gauge theory and the gauge transformation is
$\Psi \to \Psi + Q \Lambda+\cdots$.
 Symmetries of a
background $\CC$ are therefore given by solutions
to $Q \Lambda=0$. Dividing by transformations which
act trivially we see that the symetries of a background
$\CC$ may be identified with a cohomology space.
Classically, $\Psi$ has ghost
number 2 so the ghost number of $\Lambda$ is one.
Thus the symmetries of a background are identified
with the ghostnumber 1 BRST
cohomology $H^{G=1}_{BRST}(\CC)$. It can be shown that this
cohomology space has a natural structure of a Lie algebra.
Indeed, that is a corollary of the full BV/Gerstenhaber
algebra structure on $H^*$
\cite{getzler,lziii,graeme}.

{\it Examples}:

1. Consider the standard bosonic string
background $R^{1,25}$. The BRST cohomology is
$H^1=R^{26}\oplus R^{26}$, corresponding to
translations and ``dual translations.''
(The full Poincar\'e invariance of this
background arises from outer automorphisms of
the CFT. It would be interesting to study examples
of such outer automorphisms in other backgrounds.)

2. For certain string backgrounds in
$R^2$ one finds that $H^1$ is the Lie algebra of volume
preserving diffeomorphisms of a 3-dimensional cone \cite{wittzwie}.

Comparing these two examples we see that Minkowksi space is
{\it extremely unsymmetric}. These remarks lead one to ask if
there are other extremely symmetric backgrounds in string theory.
In this note we show that there are.

\vglue 0.2cm
{\elevenit \noindent 1.2. Method}
\vglue 0.1cm
We will study symmetric backgrounds in the context of the
standard 26-dimensional bosonic string. The only new point
is that we toroidally compactify {\it all} dimensions,
including time. This might sound  completely crazy,
so let us offer three justifications for doing this.

1.) As is well-known, in toroidal compactifications one
has the phenomenon of enhanced symmetry points (ESP's) where
there are length two right- or left- moving vectors in
the Narain lattice. At these points one has an enhanced
gauge symmetry associated with a $k=1$ WZW model. As we
compactify more dimensions the enhanced gauge symmetries
get bigger. Therefore, if we are searching for large symmetries
we should take this to its logical conclusion and compactify
all dimensions.

2.) One essential ingredient for the existence of the
$W_\infty$ symmetry of 2d string backgrounds is the
existence of a ``timelike CFT'' i.e., a CFT with
negative dimension vertex operators. In the 2d string
this role is played by the Liouville theory. Although
the Liouville field is a Euclidean signature boson it is
quite natural to consider {\it real} exponentials of
this field, and one has conformal dimensions
$\Delta(e^{\alpha \phi})=1-{1\over 2}(\alpha-\sqrt{2})^2$ which are
unbounded below. This example strongly suggests that if
we are to understand the fundamental symmetry principles of
string theory we will have to do peculiar things with
the time coordinate. Compactification is the simplest
solvable possibility.

3.) In the spacetime interpretation of toroidal compactification
coordinates identified by $X^i \sim X^i + 2\sqrt{2}\pi R_i$
have a breathing mode scalar in spacetime $\Phi_i(y)$
($y$=coordinates of uncompactified spacetime)
The zero-mode of this field corresponds to the
vertex operator $\p X^i \bar{\p} X^i$.
The radii may be identified with the scalar VEV's:
$R_i\sim \langle \Phi_i\rangle$. Thus $\langle \Phi_t\rangle<\infty$
might be unphysical but symmetric. We feel this is analogous
to $\langle \phi\rangle=0$ in the Weinberg-Salam model:
the background is unphysical, but better suited to
discovering the symmetries of the theory.

\vglue 0.2cm
{\elevenit \noindent 1.3. Summary of Results}
\vglue 0.1cm

\def \CB {{\cal B}}
\def \CN {{\cal N}}
\def \CL {{\cal L}}
Let us give a brief, nontechnical summary of our
results. In the remaining sections of the paper
we will be more precise about details.

First, there are some surprises in the
generalization of the standard discussion of
Narain compactification. The space of backgrounds
$\CB$ may be identified with the space of matrices
$E$ which have a decomposition into symmetric and
antisymmetric parts $E=G+B$ such that $G$ is a
quadratic form of signature $\{-1, +1^{25}\}$.
The first surprise is that $\CB$ is {\it not}
isomorphic to the homogeneous space
$O(1,25)\times O(25,1)\backslash O(26,26)$, but is only
an open proper subset. This means that the right action
by the duality group  ($=O(26,26;Z)$, with respect to an
appropriate quadratic form)
is not always defined. A much more important
new feature is that the action of the duality group turns
out to be {\it ergodic}. This means that the only
group-invariant sets are of measure zero or of total measure.
Moreover, the orbit of almost any background is dense.
It follows that the Narain moduli space, which is, roughly
speaking:
$\bigl(O(1,25)\times O(25,1)\bigr)\backslash
O(26,26)/O(26,26;Z)$ is not a manifold in the ordinary sense,
but is some kind of noncommutative manifold.

Second, the phenomenon of enhanced gauge symmetries has some
new features. In the Euclidean case there are isolated
ESP's with finite-dimensional enhanced symmetries. In the
Minkowskian case the set of ESP's is dense and the generic
enhanced gauge symmetry is infinite-dimensional.
Moreover, there is a distinguished point $E_*$ in the Narain
moduli space of toroidal compactifications. This point has
``maximal gauge symmetry'' in the sense that all states in
the BRST cohomology of the theory can be thought of as
gauge bosons of the unbroken symmetry. The name ``maximal
symmetry'' is slightly misleading. It does {\it not} mean
that if $E$ is an ESP then $H^1(E)$ has a Lie algebra
embedding in $H^1(E_*)$. One may nevertheless wish to
search for such a universal symmetry $\CL_U$, that is, a
Lie algebra which naturally includes $H^1(E)$ for all
ESP's $E$. This can be done and, curiously, involves a
Fock space corresponding formally to a 52-dimensional
``open string'' with equal numbers of space and time
directions.

Third, one may wonder how we intend to apply any of these
symmetries to spaces in which time is not compactified.
The basic idea is that at a nonsymmetric background
$E'$ we may write $E'=E_*+\Delta E$, and interpret
$\Delta E$ as a symmetry-breaking term. This corresponds
to spontaneous symmetry breaking in spacetime, since
$E\sim \langle \Phi\rangle$. We may hope to
use conformal perturbation theory to relate
correlators and Ward identities at $E_*$ to correlators
at $E'$. In particular, we propose a version of
``broken symmetry Ward identities'' below. We hope to
apply these to explain the ``high energy symmetries of
string theory'' discussed by D. Gross and E. Witten
\cite{gross,witten} in terms of the hyperbolic symmetries
of compactified time.

\vglue 0.6cm
{\elevenbf\noindent 2. Toroidal Compactifications}
\vglue 0.4cm

\vglue 0.2cm
{\elevenit \noindent 2.1. Algebraic Construction}
\vglue 0.1cm

Toroidal compactifications in $n+1$ dimensions are based on
even self-dual lattices wrt a quadratic form $D$ in
$2n+2$ dimensions. We can construct such lattices from a
generator matrix $\CE$ whose columns define basis vectors
for the lattice. Specifically we may take generator matrices:
\begin{equation}
\CM = \{\CE\in GL(2n+2;\IR):  \CE^{tr}\cdot D \cdot \CE=\check D \}
\end{equation}
where
\begin{equation}
D=\pmatrix{\eta &0\cr 0&-\eta\cr}
\qquad \check D \equiv \pmatrix{0& 1\cr 1&0\cr}
\end{equation}
 and
$\eta^{ab}_E =Diag\{1^{n+1}\}$ for Euclidean compactifications while
$\eta^{ab}_M=Diag\{-1,+1^{n}\}$ for Minkowskian compactifications.
The set $\CM$ is essentially an orthogonal group:
There is a matrix $S$ such that
$S D S = \check D $ and $S \check D S =D$ and
therefore
$\CM \cdot S =O(D,\IR)\cong S\cdot \CM=O(\check D,R)
\cong O(26,26;\IR)$.
A central theorem of the subject states that all even
unimodular lattices wrt $D$ can be obtained from such
$\CE$, so, dividing by the equivalence induced by
integral basis change we see that the moduli space of
lattices is
\begin{equation}
\IL =  \CM/O(\check D;Z)
\cong O(\check D;\IR)/O(\check D;Z)
\end{equation}
For every point $\Gamma\in \IL$ we can define
a CFT with statespace:
$ \CH_\Gamma =\oplus_{(p_L;p_R)\in \Gamma}\CF_{p_L}
\otimes \bar\CF_{p_R}$, where $\CF_p$ is the Fockspace
for the leftmovers with momentum $p$.
For discussions of universal
symmetry it is quite useful to use the notation
preferred by mathematicians: $\CH_\Gamma=
S(h_L^-)\otimes S(h_R^-)\otimes C[\Gamma]$ where
$S$ denotes the symmetric algebra and the last
factor is the group algebra of the lattice.
Identifying CFT's by left and right Lorentz transformations,
which do not change conformal weights or correlators, we
deduce that the Narain moduli space of toroidally
compactified CFT's is the double-coset:
\begin{equation}
\CN= \bigl(O(\eta)\times O(\eta)\bigr)\backslash
\CM /O(\check{D};Z)
\end{equation}

\vglue 0.2cm
{\elevenit \noindent 2.2. Sigma-model Construction}
\vglue 0.1cm

In the sigma-model approach we begin with the space of
toroidal backgrounds:
\begin{equation}
 \CB =\{ E | E=G+B, signature(G)=\eta \}
\end{equation}
and form the action:
\begin{equation}
S={1\over 2 \pi} \int_0^{2 \pi} d \sigma \int d \tau
\p X^\mu E_{\mu \nu}\pb X^\nu \qquad X^\mu\sim X^\mu+2\sqrt{2} \pi
\end{equation}
which may be quantized in the standard way. Choosing
left and right vielbeins for the metric $G$ and defining
\def \pb{\bar{\partial}}
$\p Y_L = e_L \p X ,\quad \pb Y_R=e_R \pb X$
we find the mode expansions
$ i \p Y^a_L(z)=\sum \beta_n^a z^{-n-1},
i \pb Y^a_R(z)=\sum \bar\beta_n^a \bar z^{-n-1}$ define
canonically normalized Heisenberg algebras
$[\beta^a_n,\beta^b_m] =\eta^{ab} n \delta_{n+m,0},
[\bar\beta^a_n,\bar\beta^b_m] =\eta^{ab} n \delta_{n+m,0}$.
The lattice of zero modes is obtained from the generator
matrix
\begin{equation}
\CE(e_L,e_R,E)\equiv {1\over\sqrt{2}}\pmatrix{ e_L^{a \mu}&
e_L^{a \mu} E_{\mu \nu}\cr
e_R^{a \mu}& - e_R^{a \mu} E^{tr}_{\mu \nu}\cr}
\end{equation}
and an elementary calculation shows that $\CE\in\CM$.

\vglue 0.2cm
{\elevenit \noindent 2.3. New Features of Compactified Time}
\vglue 0.1cm

The relation between the formulations can be summarized
in the following diagram:
\begin{equation}
\matrix{ &  & \CM_{\sigma} & \hookrightarrow & &\CM &  &\cr
  &\nearrow&    &   & \swarrow &  &\searrow & \cr
  \CB &  & {\buildrel \psi\over\longrightarrow }&\IH& & & &\IL\cr
         &        &        &   &\searrow& &\swarrow& \cr
         &         &     & & & \CN & &\cr}
\end{equation}
The passage to moduli space through $\IL$ is the
algebraic construction, the passage to moduli
space through $\IH\equiv O(\eta)\times O(\eta)\backslash \CM$
is the sigma-model construction.
$\CM_\sigma$ is the space of matrices of the form
in eq. (7).
The map $\CB\to \CM_\sigma$ defined by $\CE$
requires making a choice of vielbeins, so we should
really speak of the well-defined map $\psi$.

In the Euclidean case one can show that every matrix
$\CE\in\CM$ is of the form of eq. (7),
thus identifiying $\CB\cong \IH$. In the Minkowskian case this is not
true. Strictly speaking, there are models which cannot
be obtained from the sigma-model approach.

\noindent
{\it Example}: Consider compactification of $1+1$ dimensions. Let
\begin{equation}
G=\pmatrix{0&1\cr 1&0\cr} \qquad B=\pmatrix{0&-1\cr 1&0\cr}\qquad
E=\pmatrix{0&0\cr 2&0\cr}
\end{equation}
Then $E\to E^{-1}$ does not make sense. Of course, the
duality action on $\IH$ is well-defined -- it is just the
right action of $O(\check D;Z)$ on $\IH$. The corresponding
Mobius action on $E$ does not make sense because duality
takes us outside the ``sigma-model set.''

The fact that duality does not always act on
$\CB$ is not
a very serious problem because there is always some
duality transform which maps any point into the $\sigma$-model
set $\CM_\sigma$. A much more serious observation is that the
action of duality on $\IH$ and of left and right
Lorentz transformations on $\IL$ is ergodic.
Applying recently-discovered mathematical results
concerning the action of noncompact groups on
arithmetic quotients (like $\IL$) one can deduce that
for almost every point in $\IL$ the orbit under
left or right Lorentz transformations is {\it dense}.
Equivalently, for almost any background in $\IH$ the
orbit of duality is dense.

\noindent
{\it Example}: Consider again $1+1$-dimensional
compactifications. Using the
isomorphism of $o(4)$ with $sl(2)\times sl(2)$
the moduli space of lattices can be written as
\begin{equation}
\IL =\biggl[SL(2,R)/SL(2,Z) \times SL(2,R)/SL(2,Z)\biggr]/(Z_2\times Z_2)
\end{equation}
The arithmetic quotient $SL(2,R)/SL(2,Z)$ is the unit
tangent bundle (in the Poincar\'e metric) over the modular
curve $\Sigma=SO(2)\backslash SL(2,R)/SL(2,Z)$. In Euclidean
compactifications we must divide $\IL$ by $SO(2)\times SO(2)$.
The left-action of $SO(2)$ on $SL(2,R)/SL(2,Z)$
is simply rotation in the fibers.  In the Minkowskian case
we must divide $\IL$ by $SO(1,1)\times SO(1,1)$ .
The left-action of $SO(1,1)$ on $SL(2,R)/SL(2,Z)$
is left-multiplication by
\begin{equation}
\pmatrix{\cosh t&\sinh t\cr \sinh t& \cosh t\cr}
\end{equation}
This action is simply geodesic flow in the unit
tangent bundle. The
typical geodesic on $\Sigma$ is dense. Thus
we see that the moduli space of conformal field theories
is a nonstandard space. It is a typical example of
a noncommutative manifold.

\vglue 0.5cm
{\elevenbf \noindent 3. Enhanced Gauge Symmetries}
\vglue 0.4cm

\vglue 0.2cm
{\elevenit \noindent 3.1. Enhanced Symmetry Points}
\vglue 0.1cm

Recall that the unbroken symmetry at a background is
the ghost number one cohomology. Since $\CH_\Gamma$ is
a sum of Fock spaces this is easily calculated:
\begin{equation}
H^1(\CH_\Gamma)=
\oplus_{p_L,p_R} \Biggl[H^0(\CF_{p_L})\otimes H^1(\bar\CF_{p_R})
\oplus H^1(\CF_{p_L})\otimes H^0(\bar\CF_{p_R})\Biggr]
\end{equation}
Since $H^*(\CF_p)=0$ for $p^2\notin \{2,0,-2,-4,\dots\}$
we see that for generic points in $\IL$ $H^1=R^{26}\oplus R^{26}$:
the generic background is just as symmetric as Minkowski space.
Let us define a background
$\Gamma\in \IL$ to be an enhanced symmetry point (ESP) if
 $\exists$ $(p_L;0)\in\Gamma$
or $(0;p_R)\in\Gamma$ with $p_L^2$ or $p_R^2$
$\in \{2,0,-2,-4,\dots\}$

There are infinite-dimensional unbroken
gauge symmetries at a dense set of points in
$\IL$. This is elementary:
Consider lattices obtained from $\CE$ such
that $\sqrt{2}\CE$ is a rational matrix. Moreover, for
the generic such point the purely left- and right-moving
sublattices
$(\gamma_L;0)\oplus (0;\gamma_R)\subset \Gamma$
will have maximal rank. In particular they will have
an infinite number of points inside the forward and
backward light cone and therefore the unroken symmetry algebras
are typically infinite-dimensional. The unbroken
symmetries belong to a class of Lie algebras known
as ``generalized Kac-Moody Lie algebras'' \cite{borcherds}.
The importance of these algebras in string theory has
been emphasized by Goddard and Olive \cite{godol}. We will
refer to these infinite-dimensional gauge symmetries
as ``hyperbolic symmetries.''
\footnote{
\ninerm\baselineskip=11pt In general these
Lie algebras are not
hyperbolic Kac-Moody algebras.
}

\vglue 0.2cm
{\elevenit \noindent 3.2. A Distinguished Compactification}
\vglue 0.1cm

One interesting new feature of timelike compactification
is that there is a distinguished point in the moduli space
of CFT's:

{\bf Proposition 1}: $\exists$! $\Gamma_*\in \CN$ for which
\begin{equation}
\CH_{\rm closed}= \CH_{\rm open}\otimes \bar\CH_{\rm open}
\end{equation}

The proof of this is trivial. The string factorizes iff
the lattice of zero modes is a direct sum of left- and
right-moving lattices. This implies that the left- and
right-moving sublattices are even unimodular. Since the
signature of these lattices is $(1,25)$ they are unique.
Therefore $\CH_{\Gamma_*}=\CC \otimes \bar{\CC}$
where $\CC=S(h_{1,25}^-)\otimes C[II^{1,25}]$.

Several remarks are in order. The
symmetry $H^1(\CC)$ is in a sense maximal since the
only representation which appears in the physical spectrum
is the adjoint representation $H^{0,1}\otimes H^{1,0}$
for zero modes of ``gauge bosons.'' These are states
whose existence is dictated by the existence of
symmetry. Moreover the form of the string densities
is completely fixed by the unbroken symmetry
(although the integrals over moduli space typically
diverge) \cite{finite}. This partially realizes the
idea that unbroken fundamental symmetries of strings
completely fix the string $S$-matrix.

The unbroken symmetry
algebra $H^1(\CC)$ is related to the Monster group and
has been studied in this context in \cite{borcherds}.
Given the remarks in the previous section
one might wonder if this ``Monster orbit'' is dense;
one can show that this is not the case.

\vglue 0.2cm
{\elevenit \noindent 3.3. Universal Symmetry}
\vglue 0.1cm

Let us define a ``universal symmetry for toroidal
compactifications'' to be a
Lie algebra $\CL$ that contains $H^1(\CH_\Gamma)$
for all ESP's $\Gamma$. There is not a unique
choice for such a symmetry, but we would like to find one
which is natural and, in some sense, minimal.

Given the distinguished symmetry of the previous
subsection one might expect that $H^1(\CH_{\Gamma_*})$
is a universal symmetry. This is not true.
However, there is
a very simple construction which does give such
a symmetry. First we must put left-movers and
right-movers together into a single multiplet:

\begin{eqnarray}
&\rho_n^A = \beta_n^A  \qquad A=1,\dots 26 &
\nonumber \\
&\rho_n^A  = \bar{\beta}_{-n}^{A-26}  \qquad A=27,\dots 52 &
\nonumber \\
&[\rho_n^A,\rho_m^B] =D^{AB}n \delta_{n+m,0} &
\end{eqnarray}
Let $h_{26,26}$ be the corresponding Heisenberg algebra.
The commutator equation is clearly Lorentz invariant.
We now introduce $\Gamma^{26,26}$, the
even unimodular lattice in
$52$ dimensions, of signature $(26,26)$.
It is necessary to consider a larger lattice
$\tilde \Gamma^{26,26}$ in $C^{52}$ obtained by allowing
integer combinations of purely real or imaginary vectors.
Finally we have

\noindent
{\bf Proposition 2}: Let
\begin{equation}
\tilde\CH=S(h^-_{26,26})\otimes C[\tilde{\Gamma}^{26,26}]
\end{equation}
Then the Lie algebra:
\begin{equation}
\CL_U\equiv
\tilde\CH[1]^{Vir^+}/(\tilde\CH[1]^{Vir^+}\cap Vir^-\cdot \tilde\CH)
\end{equation}
is a universal symmetry. Here $\tilde\CH[1]^{Vir^+}$
denotes restriction to the Virasoro dimension $=1$ primaries.

This
construction is closely related to some work of
Giveon and Porrati \cite{giveon}

\vglue 0.5cm
{\elevenbf \noindent 4. Symmetry-Breaking \hfil}
\vglue 0.4cm
Physically, time is not compact. Nevertheless, we
may try to describe backgrounds with noncompact time as
broken symmetry phases of the hyperbolic symmetries
of compactified time. The paradigm for this description
is the compactification of a single Euclidean coordinate on
a circle of radius $R$. At the self-dual radius $R=1$
there is an enhanced affine $su(2)_L^{(1)}\times su(2)_R^{(1)}$
symmetry in the conformal field theory. This leads to
a spacetime $su(2)\times su(2)$ gauge symmetry. If
we increase $R$ to $R>1$ then, from the point of view of
the spacetime theory, the gauge symmetry is spontaneously
broken by the vev of the scalar field $\Phi_R(x)$ whose
zero-mode corresponds to the modulus of the circle
$\Phi_R(p=0)\leftrightarrow \p X \pb X$. One way to try to
deduce quantitative consequences of this point of view is
to write down the broken Ward identities for the broken
symmetries, analogous to the Slavnov-Taylor identities
used in the proofs of renormalizability of spontaneously
broken gauge theories. Our goal for the remainder of this
note will be to write down a set of broken symmetry Ward
identities. The results of this section apply to all
toroidal compactifications, and are independent of the
signature of the compactified dimensions.

The basic strategy is quite simple and best explained in
the sigma model picture. If $E$ is an ESP
and $E'=E+\Delta E$ is not then we may still relate
correlation functions in one theory to those in the
other by conformal perturbation theory. Since the
action is linear in $E$ this appears to be completely
trivial: we simply write
\begin{eqnarray}
&\langle\prod_i V_i \rangle_{E'} =\int
e^{-{1\over 2 \pi} \int \p X E' \pb X}\prod_i V_i &
\nonumber \\
&=\langle e^{-{1\over 2 \pi} \int \p X \Delta E \pb X}
\prod_i V_i\rangle_E \qquad \qquad .&
\end{eqnarray}

Of course, eq. (17) is much too naive. Conceptually it does not
make sense: In CFT we identify states and operators, but
the state-spaces
of the two conformal field theories $\CH_E,\CH_{E'}$
are different. For example, they are built from different
representations of the Virasoro algebra.
At best we can hope that conformal
perturbation theory will define a linear map
$T^{E',E}:\CH_E\to \CH_{E'}$. The  manipulations in eq. (17) are
also technically wrong because, if we attempt to expand the
interaction we will obtain infinity for the integrals of
the singular correlators.

\vglue 0.2cm
{\elevenit \noindent 4.2. Lorentz Transport}
\vglue 0.1cm

In \cite{finite} the conformal perturbation series
(for toroidal compactifications) is given a precise
meaning and the series is summed up to produce a
globally well-defined transport on the moduli space
of toroidally compactified theories. To state the
result we must introduce some notation.

Recall that $E\in \CB$, and via
$\psi:\CB\hookrightarrow\IH=(O(\eta)\times O(\eta))\backslash \CM$
we can define a right-action of $O(\check D,R)$ on $\CB$.
(This may be written explicitly as a matrix M\"obius
transformation on $E$.) If we
block-decompose the matrix $g\in O(D)$ as:
\begin{equation}
g=\pmatrix{g_{11}& g_{12}\cr g_{21} & g_{22}\cr}
\end{equation}
and define natural coordinates on $\IH$:
\begin{equation}
\Delta \CE(g)
= -g_{21}^{tr} (g_{22}^{tr})^{-1}\eta=-\eta (g_{11})^{-1}g_{12}
\end{equation}
we can then state:

\noindent
{\bf Proposition 3}: Let $E'=E\cdot g$. The integrals
in conformal perturbation theory can be defined so that:
\begin{equation}
{\bigl\langle e^{-{1\over 2 \pi}\int \p Y\cdot \Delta \CE\cdot \pb Y}
\prod V_i\bigr\rangle_E\over
\bigl\langle e^{-{1\over 2 \pi}\int \p Y\cdot \Delta \CE\cdot \pb Y}
\bigr\rangle_E} = \bigl\langle \prod T^g(V_i)\bigr\rangle_{E'}
\end{equation}
where the transport
\begin{equation}
T^g:\CH_E \longrightarrow \CH_{E'}
\end{equation}
is defined by
\begin{eqnarray}
&T^{g}: |p_L;p_R\rangle \to |g\cdot(p_L;p_R)\rangle &
\nonumber \\
& \rho_n^A\to \rho_n^{A'} (D g D)_{A'}^{\ A} &
\end{eqnarray}

Remark: The operator $T^g$ defines a parallel transport on
the ``bundle of conformal field theories'' $\CH\to\IL$,
or $\CH\to\IH$. The relation to the Zamolodchikov metric
is provided by the projection $\Pi$ to the space of
exactly marginal operators which may be identified with
$T \IH$. One can show that
\begin{equation}
\bigl\langle \Pi  T^g(\beta_{-1}^a \bar{\beta}_{-1}^{\bar{a}})(1)
\Pi T^g(\beta_{-1}^b \bar{\beta}_{-1}^{\bar{b}})(0)\bigr\rangle
d(\Delta \CE_{a \bar{a}}) d(\Delta \CE_{b \bar{b}})
\end{equation}
is the right-invariant Zamolodchikov metric
$Tr[G^{-1} d E G^{-1} dE^{\rm tr}]$.

These remarks are related to the works
\cite{rangan,rsz} and to B. Zwiebach's contribution to
this volume.

\vglue 0.2cm
{\elevenit \noindent 4.3. Broken Ward Identities}
\vglue 0.1cm

Suppose we are at an enhanced symmetry point $E$, and
suppose $J$ is a BRST invariant current. Then the Ward
identities associated with $J$ are:
\begin{equation}
0=\sum_i \langle \oint_{z_i} J(z)\cdot V^{i}
\prod_{j\not= i} V^{j}\rangle_E
\end{equation}
Now consider another background $E'=E\cdot g=E+\Delta E$
where the symmetry is broken. This will happen if the
conformal perturbation
$\CO=\p Y\cdot \Delta \CE\cdot \pb Y$ does not commute with
the symmetry, that is, if
\begin{equation}
\delta_J \CO(z,\bar z)=\oint_z J(w)dw \CO(z,\bar z)\not= 0
\end{equation}

\noindent
{\bf Proposition 4}: The broken Ward identities are
\begin{eqnarray}
&0 = \int_{|w-z_i|> \epsilon} d^2 w \langle
T^{g}(\delta_J\CO) \prod T^{g} V^{i}(z_i)\rangle_{E'} + &
\nonumber \\
&\sum_i \langle \biggl[\oint_{|w-z_i|=\epsilon}
dw_\mu (T^{g} J^\mu)(w,\bar{w})
T^{g} V^{i})(z_i,\bar z_i)\biggr] \prod_{j\not= i}
(T^{g}V^{j})\rangle_{E'} &
\end{eqnarray}

The proof of this formula is based on an identity
\begin{equation}
T^g(\delta_J \CO)(z,\bar{z}) = - {\p \over \p \bar{z}} (T^g J(z,\bar{z})) -
{\p\over \p z}(T^g\bar J(z,\bar{z}))
\end{equation}
which defines a current $J^\mu=(J,\bar{J})$.
The component $\bar J$ may be constructed explicitly
from $J$ and $g$
\cite{finite}. The broken Ward identity can be given the
following physical interpretation. The vertex operator
$T^g(\delta_J \CO)(z,\bar{z})$ is the vertex operator for
the emission of a zero-momentum goldstone boson $\pi_J(0)$
for the broken symmetry $J$. ($\pi_J$ gets eaten by the vector
bosons.)
Define $\delta V_i$ by the operator appearing
with a simple pole in the ope of $TJ^\mu$ with $V_i$. Then
proposition 4 implies an equality of analytically continued
$S$-matrix elements $\CA$
\begin{equation}
\CA(\pi_J(0),V_i)=\sum_i \CA(V_1,\dots \delta V_i,\dots V_n)
\end{equation}
Which may be regarded as a ``low-energy theorem.'' Thus the
same identity can be simultaneously regarded as a statement
of explicit symmetry-breaking on the worldsheet and spontaneous
symmetry-breaking in spacetime.

\vglue 0.5cm
{\elevenbf \noindent 5. Conclusion \hfil}
\vglue 0.4cm

We conclude with two speculations. First, we believe that
the broken Ward identities can be used to explain the
high energy symmetries of string theory which were the
subject of some speculation 5 years ago \cite{gross,witten}.
We hope this will result from a combination of the results
of sections two and four.
Second, we think that it is significant that the
construction of a universal symmetry for toroidal
compactifications involves the ``52-dimensional open string.''
Many facts about closed string field theory strongly
suggest that it is a broken symmetry phase of ....
Perhaps it is a broken symmetry phase of some such
``open string.''

\vglue 0.5cm
{\elevenbf\noindent 6. Acknowledgements \hfil}
\vglue 0.4cm

I would like to thank the Ecole Normale Superieure for
hospitality while this note was written. This work is
also supported by DOE grant DE-AC02-76ER03075,
DOE grant DE-FG02-92ER25121,
and by a Presidential Young Investigator Award.

\vglue 0.5cm
{\elevenbf \noindent 7. References \hfil}
\vglue 0.4cm

\bibliographystyle{plain}

\end{document}